\documentclass[conference]{IEEEtran}

\usepackage{cite}
\usepackage{booktabs}
\usepackage{amsmath}
\usepackage{graphicx}
\usepackage{multirow}
\usepackage{array}
\usepackage{enumitem}
\usepackage{xcolor}
\usepackage{url}
\usepackage{balance}
\usepackage{tabularx}
\usepackage{float}
\usepackage{listings}

\usepackage{pifont}
\usepackage{tcolorbox}

\usepackage[normalem]{ulem}

\newcolumntype{Y}{>{\raggedright\arraybackslash}X}

\floatstyle{ruled}
\newfloat{algorithm}{t}{loa}
\floatname{algorithm}{Algorithm}

\title{From Natural Language to Verified C: Verification-Guided Generation with ACSL Specifications}

\author{
\begin{tabular}{@{}c@{\hspace{2.5em}}c@{}}
\begin{minipage}[t]{0.30\textwidth}
\centering
{\normalfont\sublargesize Zian Wang\par}
{\normalfont\normalsize East China Normal University\newline
Shanghai, China\newline
51275902168@stu.ecnu.edu.cn\par}
\end{minipage}
&
\begin{minipage}[t]{0.30\textwidth}
\centering
{\normalfont\sublargesize Qin Li\par}
{\normalfont\normalsize East China Normal University\newline
Shanghai, China\newline
qli@sei.ecnu.edu.cn\par}
\end{minipage}
\\[3ex]
\begin{minipage}[t]{0.30\textwidth}
\centering
\vspace*{2ex}
{\normalfont\sublargesize Weisong Sun\par}
{\normalfont\normalsize Nanyang Technological University\newline
Singapore\newline
weisong.sun@ntu.edu.sg\par}
\end{minipage}
&
\begin{minipage}[t]{0.30\textwidth}
\centering
\vspace*{2ex}
{\normalfont\sublargesize Haibin Cai\par}
{\normalfont\normalsize East China Normal University\newline
Shanghai, China\newline
hbcai@sei.ecnu.edu.cn\par}
\end{minipage}
\end{tabular}
}

\begin{document}
\maketitle

\begin{abstract}
AI-assisted code generation amplifies the demand for automated formal verification. Generated programs must not only be executable but also accompanied by machine-checkable evidence of correctness. 
Existing approaches can synthesize functionally plausible code or plausible-looking specifications from natural-language prompts; however, they do not explicitly co-design the implementation and specification as a unified artifact optimized for verification. 
Consequently, automated verifiers often fail to discharge proof obligations. Typical failure modes include missing or implicit proof anchors, program structures that are hostile to automated reasoning, and inconsistencies between code-level state and specification-level requirement facts arising from the independent generation of code and specifications. Even when verification succeeds, it may only establish weak specifications that omit requirement-critical behavior.
This paper introduces VeriGen-C, a formal-verification-friendly framework for the joint synthesis of C programs and ACSL specifications from natural-language requirements. 
Code generation and repair are constrained to expose proof-relevant state, to streamline control flow for deductive verification, and to eliminate implementation patterns that hinder automated proof search. Specification generation and refinement are similarly constrained to mitigate verification timeouts and to ensure alignment with the user requirements. 
Empirical evaluations on HumanEval-X, FM-Bench, and LiveCodeBench Lite demonstrate that VeriGen-C substantially improves the rate of verified and requirement-aligned solutions relative to an interface-compatible baseline, from 5.5\% to 44.5\%, from 44.3\% to 80.0\%, and from 6.5\% to 58.7\%, respectively.
\end{abstract}


\section{Introduction}
\label{sec:introduction}

AI-assisted code generation has become pervasive in software development. Recent industrial reports document widespread developer adoption and extensive deployment of automatically generated code in production workflows~\cite{stackoverflow2025devsurvey_ai, alphabet2024q3_earnings_call}. Within this context, assurance increasingly emerges as the principal bottleneck. For C programs, testing and manual code review remain indispensable; yet, exhaustive coverage of boundary conditions, memory accesses, arithmetic constraints, and loop-dependent behavior is rarely practical~\cite{myers2011art,councilman2025formalverificationllmgeneratedcode}.

Formal verification offers a mathematically rigorous approach to increase confidence in generated code by validating program behavior against explicit specifications. For C programs, the verification target is not the executable implementation alone, but a composite C/ACSL artifact. Its source code, contracts, frame conditions, and loop invariants must all be syntactically correct, semantically aligned with the requirement, and amenable to automated reasoning~\cite{acsl_manual}. If either the implementation or the specification is generated incorrectly, the resulting artifact may become unverifiable, or may be verified only with respect to a specification that is too weak to capture the intended requirement. Moreover, AI-generated coding patterns can introduce redundant or proof-irrelevant structures, which in turn induce superfluous verification obligations and degrade tool performance~\cite{li2026performance_ai_code}.

It is therefore crucial to synthesize artifacts whose implementations and specifications are both suitable for automated verification. Example~\ref{lst:intro_find_first_contrast} illustrates this point via a simple requirement: given an array, return the first index whose value equals \texttt{x}, or \texttt{-1} if no such index exists. The example contrasts verification-friendly and verification-unfriendly realizations.

\begin{samepage}
\begin{lstlisting}[language=C,basicstyle=\ttfamily\scriptsize,commentstyle=\ttfamily\scriptsize,linewidth=0.98\columnwidth,xleftmargin=0.03\columnwidth,xrightmargin=0pt,framexleftmargin=0.5mm,framexrightmargin=0.5mm,aboveskip=2pt,belowskip=2pt,caption={Verification-Friendly vs. Verification-Unfriendly Artifacts},label={lst:intro_find_first_contrast}]
/* Requirement: first index k with a[k] == x, or -1. */

/* (*@\textbf{(A) Implementation issue | WP: Fail\_10\_19}@*) */
/*@
  requires n >= 0 && (n == 0 || \valid_read(a + (0 .. n-1)));
  ... same first-index postconditions as below ...
*/
int find_first_ptr(const int *a, int n, int x) {
  const int *p = a, *end = a + n;
  while (p < end) {
    if (*p == x) return (int)(p - a);
    p++;
  }
  return -1;
}
/* (*@\textbf{(B) Missing bridge invariant | WP: Fail\_21\_22}@*) */
/*@
  ...
  ensures \result >= 0 ==>
    \forall integer k; 0 <= k < n ==> (a[k] == x ==> \result <= k);
*/
int find_first_heavy_spec(const int *a, int n, int x) {
  int i = 0, res = -1;
  /* invariants track res and 0..i, but omit:
     res >= 0 ==> all k < res have a[k] != x. */
  while (i < n) { ... }
  return res;
}
/* (*@\textbf{(C) Coupled artifact | WP: Pass\_24\_24}@*) */
/*@
  requires n >= 0 && (n == 0 || \valid_read(a + (0 .. n-1)));
  ensures \result == -1 ==>
    \forall integer k; 0 <= k < n ==> a[k] != x;
  ensures \result >= 0 ==> a[\result] == x;
  ensures \result >= 0 ==>
    \forall integer k; 0 <= k < \result ==> a[k] != x;
*/
int find_first_friendly(const int *a, int n, int x) {
  int i = 0, res = -1;
  /* invariants connect res with processed prefix 0..i. */
  while (i < n) {
    if (res == -1 && a[i] == x) res = i;
    i++;
  }
  return res;
}
\end{lstlisting}
\end{samepage}

In artifact (A) of Example~\ref{lst:intro_find_first_contrast}, \texttt{find\_first\_ptr} conceals the searched prefix behind pointer offsets and an early return. Under Frama-C/WP~\cite{kirchner2015framac,framac_wp_manual}, this control and data-flow structure results in a verification-time timeout. The proof attempt must reconstruct an indexed-prefix property from pointer arithmetic \texttt{p - a} and from a loop exit that occurs within the search body, which provides no stable loop-level anchor characterizing the already-processed segment of the array. Artifact (B), \texttt{find\_first\_heavy\_spec}, employs an index-based loop and a semantically rich first-occurrence postcondition. However, it omits the invariant \(\texttt{res} \ge 0 \Rightarrow \forall k < \texttt{res}.\ a[k] \ne x\), so the verifier cannot derive the firstness postcondition \(\forall k.\ a[k]=x \Rightarrow \texttt{res}\le k\) from the available loop facts. Artifact (C), \texttt{find\_first\_friendly}, simultaneously restructures the implementation and the specification: the implementation explicitly exposes the processed-prefix state, and the invariants and postconditions express firstness in terms of this shared anchor, enabling the verifier to discharge all proof obligations. This example highlights a more general synthesis and generation challenge: implementations must expose proof-relevant state, specifications must supply the bridging properties required by the verifier, and both must be co-designed around common proof anchors. Existing requirement-formalization and specification-generation approaches~\cite{giannakopoulou2020fret,cosler2023nl2spec,ma2025req2ltl,ma2025specgen,wen2024autospec,SU2026115177} address important aspects of this problem, but they typically formalize natural-language requirements or synthesize specifications for a given program, rather than jointly constructing C code and ACSL specifications from the same natural-language description.

The same coupling challenge persists in the aftermath of verification failure. A non-passing weakest-precondition (WP) outcome may stem from ACSL surface form, missing proof bridges, or implementation structure. Treating these failure modes as independent regeneration targets can disrupt the alignment between program state and contract clauses. Verifier-in-the-loop systems have shown the value of proof feedback in Dafny, Verus, and broader formal-methods workflows~\cite{loughridge2024dafnybench,yang2025autoverus,wang2026preguss,ding2026fmagent,ma2026specsyn,li2026osvbench,cheng2026sysmobench}. Nevertheless, code and specifications risk losing shared proof anchors when they are generated or revised as separate artifacts. The implementation may employ compact control-flow structures, low-level pointer manipulations, or library-dependent operations, whereas the specification may articulate requirements in terms of witnesses, value ranges, or intermediate invariants that are never explicitly represented in the code. This divergence forces the verifier to reconstruct implicit relationships between program state and contract clauses. In practice, it frequently yields either undischarged proof obligations or specifications that are progressively weakened until they no longer capture the intended behavior.

This paper proposes VeriGen-C, a formal-verification-friendly framework for joint generation of C programs and ACSL specifications from natural-language requirements. VeriGen-C produces verifier-aware code and ACSL annotations as a tightly coupled artifact. It analyzes non-passing WP outcomes within a bounded repair loop and directs repair actions between specification-only refinement and conservative joint revision of both code and specification. Finally, it performs a requirement-alignment check, so reported success requires both implementation–specification conformance and non-vacuous specification content.

VeriGen-C connects verifier-guided repair, specification synthesis, and verification-aware prompting through a shared proof-anchor interface: the implementation exposes named state, the contract refers to it, and repair preserves that relation. Unlike SpecGen and AutoSpec, which start from an existing program, VeriGen-C decides implementation shape and contract together. A clause-level alignment filter after WP success ensures that verified contracts retain task-specific content.

\noindent\textbf{Contributions.}
This work makes the following contributions.
\begin{itemize}
    \item We propose VeriGen-C, a joint code-specification generation framework for verification-friendly C/ACSL artifacts. Unlike approaches that synthesize code and specifications as separate outputs, VeriGen-C constructs the C code and ACSL annotations around shared proof anchors. These anchors make program state, loop invariants, and function contracts support the same verification objectives.
    
    \item We introduce verification-oriented construction and refinement principles for generated C/ACSL artifacts. These principles govern both initial synthesis and repair. Implementations are shaped to expose proof-relevant state; ACSL clauses are structured to capture requirement-relevant behavior using verifier-supported forms; and refinement steps are constrained to preserve the code--specification coupling needed for automated proof.
    
    \item We empirically evaluate VeriGen-C on HumanEval-X, FM-Bench, and LiveCodeBench Lite. Compared with an interface-compatible one-shot baseline, VeriGen-C improves WP-verified, requirement-aligned success from 5.5\% to 44.5\%, from 44.3\% to 80.0\%, and from 6.5\% to 58.7\%, respectively, without manual post-editing.
\end{itemize}

\section{Background and Related Work}
\label{sec:background_related_work}

\subsection{C/ACSL Verification Background}
\label{subsec:verification_background}

ACSL is a behavioral specification language for C programs that supports function contracts, loop invariants, frame conditions, termination metrics (variants), and assertions expressed as source-code annotations~\cite{acsl_manual}. Frama-C/WP verifies annotated C programs by means of weakest-precondition calculus: it computes verification conditions from a C implementation together with its ACSL annotations, and then attempts to discharge these conditions using automated theorem provers such as Alt-Ergo and Z3~\cite{framac_wp_manual,altergo_official,demoura2008z3}. A successful WP analysis establishes that the implementation satisfies the given ACSL specification under the chosen verification configuration.

The resulting proof is inherently specification-relative. In C/ACSL, the implementation must expose proof-relevant program state, including bounds, counters, witnesses, flags, and output ranges. The specification must, in turn, be closely aligned with the program’s control flow, memory footprint, and intermediate states. If the code obscures proof anchors or if the contract asserts properties that are not supported by the structure of the implementation, automated WP verification may fail even when the intended behavior is conceptually simple. Consequently, natural-language-to-verified-C synthesis constitutes a coupled code-and-specification generation problem, rather than a pipeline of independent code generation followed by subsequent annotation.

We use the term \emph{proof anchor} for a named, observable program state---bound, counter, flag, witness, buffer, or result---that both loop invariants and function contracts can reference. It is effective when the implementation exposes it and the contract states its relation to the requirement. In the first-index example (Section~\ref{sec:introduction}), \texttt{i} anchors the processed prefix and \texttt{res} the returned witness. This relation organizes construction and repair below.

\begin{table*}[!t]
    \centering
    \caption{Positioning of VeriGen-C against related artifact-generation settings. ``P'' denotes partial coverage of a dimension.}
    \label{tab:related_work_matrix}
    \footnotesize
    \setlength{\tabcolsep}{6pt}
    \begin{tabular}{@{}lccccccc@{}}
    \toprule
    Research line &
    \shortstack{NL req.\\input} &
    \shortstack{C\\impl.} &
    \shortstack{Formal\\spec.} &
    \shortstack{Joint code\\+ spec} &
    \shortstack{Tool-backed\\proof} &
    \shortstack{Verifier-guided\\repair} &
    \shortstack{Req.-content\\check} \\
    \midrule
    Requirement formalization, e.g., FRET/nl2spec & Y & N & Y & N & P & N & P \\
    Specification generation, e.g., SpecGen/AutoSpec & P & N & Y & N & Y & P & P \\
    Verifier-in-the-loop generation, e.g., DafnyBench/AutoVerus & P & P & Y & N & Y & P & N \\
    Existing-artifact scaling, e.g., Preguss/OSVBench & P & P & Y & N & Y & Y & P \\
    VeriGen-C & Y & Y & Y & Y & Y & Y & Y \\
    \bottomrule
    \end{tabular}
\end{table*}

\subsection{Problem Formulation}
\label{subsec:problem_formulation}

In natural-language-to-verified-C generation, the input is a natural-language requirement \(r\). The intended output consists of a C implementation \(c\) together with an ACSL specification \(s\). These jointly constitute an annotated program that can be statically verified using Frama-C/WP, such that the resulting verified contract preserves substantive content from the original requirement. We model the terminal output as
\begin{equation}
\mathcal{O}=(c,s,v,m),
\end{equation}
where \(v\) denotes the WP verification outcome for the pair \((c,s)\), and \(m\) is the requirement-alignment report obtained by comparing \(r\) with \(s\). The entire generation procedure is automated and operates under a bounded repair budget; given \(r\), the system executes to a terminal outcome without any human intervention on the generated C+ACSL artifact.

The reported success criterion is the conjunction
\begin{equation}
\mathsf{WP}(c,s)=\textsc{Pass}
\land
\mathsf{Accept}(\mathsf{ReqAlign}(r,s)).
\end{equation}
where \(\mathsf{WP}(c,s)\) denotes the outcome produced by Frama-C/WP, and the function \(\mathsf{Accept}(\cdot)\) maps the requirement-alignment report into the subset of accepted categories defined in Section~\ref{sec:reqalign-checking}. This criterion is strictly stronger than requiring only successful code generation or a passing WP result: the artifact must both be formally proved correct with respect to its specification, and that specification must demonstrably preserve non-trivial content from the original requirement.

\subsection{Related Work}

\textbf{Positioning.}
Table~\ref{tab:related_work_matrix} situates VeriGen-C with respect to the primary artifact it produces and the role of formal verification within its workflow. The table serves to distinguish problem formulations rather than to rank systems that operate over heterogeneous inputs, targets, and verification formalisms. The key distinction is not that VeriGen-C encompasses a greater number of stages, but that it treats the C implementation, ACSL specifications, verifier feedback, and requirement-alignment evidence as tightly coupled components of a single, integrated verified-code generation process.

\textbf{Requirement formalization.}
A substantial body of work addresses the transformation of informal requirements into structured or formal artifacts. This line of research includes consistency checking between requirements and formal specifications, translation of requirements into analyzable models, controlled natural-language frameworks such as FRET, and LLM-assisted translation of requirements into regular expressions, temporal logic, or LTL~\cite{kozlenkov2002design,ghosh2016arsenal,giannakopoulou2020fret,giannakopoulou2021automated,hahn2022formal,cosler2023nl2spec,ma2025req2ltl}. These systems strongly motivate requirement-preserving formalization, but they typically terminate at a formal statement or model and do not produce an end-to-end pipeline that yields executable C code, ACSL contracts, and corresponding Frama-C/WP proof artifacts.

\textbf{Formal specification generation.}
Formal-specification generation approaches synthesize annotations or contracts for existing programs or specified formal targets. For instance, SpecGen generates and repairs Java specifications, while AutoSpec produces C/ACSL annotations for existing C code via program analysis and validation techniques. NL2ACSL translates natural-language descriptions into ACSL contracts~\cite{ma2025specgen,flanagan2001houdini,ernst2007daikon,wen2024autospec,SU2026115177,beg2026evaluatingllmgeneratedacslannotations}. These systems are closely related to VeriGen-C with respect to the specification dimension; however, they address a different task boundary. VeriGen-C operates prior to the existence of an implementation and must co-design the C control structures, memory interfaces, proof anchors, and ACSL clauses as part of a unified synthesis process.

\textbf{Verifier-in-the-loop generation and repair.}
Recent work demonstrates that feedback from automated verifiers can effectively guide program and proof generation or repair. DafnyBench evaluates LLM-generated proof hints for Dafny~\cite{loughridge2024dafnybench}, AutoVerus repairs Rust/Verus proofs~\cite{yang2025autoverus}, and FM-Agent leverages LLM-assisted reasoning over large-scale formal-methods artifacts~\cite{ding2026fmagent}. Recent Dafny work follows the same broad principle: Laurel uses LLM-generated hints to unblock proofs~\cite{mugnier2025laurel}, and Silva et al. infer multiple helper Dafny assertions with LLMs~\cite{silva2025dafny}. These systems establish the value of verifier feedback for proof repair; VeriGen-C applies that principle to a jointly generated C/ACSL artifact, where implementation shape and contract clauses are revised together. Additional efforts focus on scaling verification and specification synthesis for existing infrastructures, including Ansible-based systems, large-scale software, operating-system verification, and formal-modeling benchmarks~\cite{councilman2025formalverificationllmgeneratedcode,wang2026preguss,ma2026specsyn,li2026osvbench,cheng2026sysmobench}.

\section{Methodology}
\label{sec:method}

\begin{figure*}[!t]
    \centering
    \includegraphics[width=\textwidth]{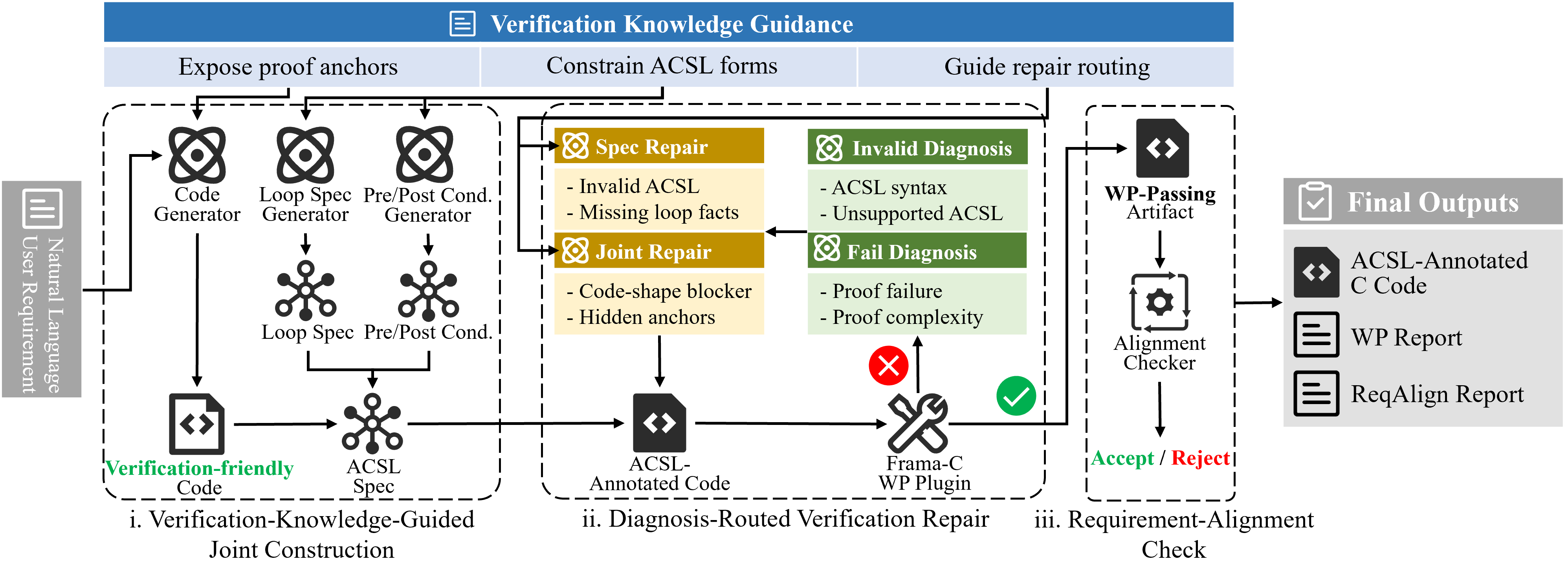}
    \caption{Overview of VeriGen-C. Verification knowledge guides joint construction of C code and ACSL specifications, diagnosis-routed repair handles non-pass Frama-C/WP outcomes, and WP-passing artifacts are filtered by requirement-alignment checking before final reporting. LLM-based generation is paired with deterministic verification, normalization, and routing.}
    \label{fig:framework}
\end{figure*}

Figure~\ref{fig:framework} illustrates the VeriGen-C workflow. Given a natural-language requirement, VeriGen-C jointly synthesizes a C/ACSL artifact under verification-knowledge constraints. The implementation exposes proof anchors, and the ACSL specification encodes requirement-relevant behavior at those anchors using Frama-C/WP-supported constructs. The resulting artifact is checked by Frama-C/WP; when verification does not pass, VeriGen-C diagnoses the failure and initiates either specification-only repair or joint code-and-specification repair. Once an artifact satisfies the WP proof obligations, it proceeds to a requirement-alignment phase. The alignment checker accepts artifacts whose verified contracts retain non-trivial requirement content and rejects artifacts classified as shell-only, inconsistent, or unclear. The workflow follows a central design premise: implementation shape and specification shape must be co-generated because both determine whether automated deductive verification succeeds.

VeriGen-C is modeled as a bounded state machine. At iteration \(i\), the current artifact is \(x_i = (c_i, s_i)\), where \(c_i\) denotes the C implementation and \(s_i\) denotes the ACSL specification. The task state \(\sigma_i\) maintains the requirement, current artifact, WP verification outcome, diagnostic information, selected repair strategy, requirement-alignment report when available, and the remaining computational budget. If the current artifact passes WP, it exits the repair loop and is subjected to the requirement-alignment check. Otherwise, VeriGen-C continues to repair the artifact until the budget is depleted or a terminal failure condition is encountered. The final output comprises the ACSL-annotated C code, the WP verification report, and the requirement-alignment accept/reject decision.

VeriGen-C separates generative decisions from deterministic control. Construction uses three LLM calls (C, loop annotations, and function contract); each repair round uses a diagnosis call and a repair call; alignment uses one clause-decomposition call. Extraction, preflight, ACSL normalization, failure parsing, route selection, and bucket mapping are deterministic. Frama-C/WP remains the proof authority.

\subsection{Verification-Knowledge-Guided Joint Construction}
\label{subsec:stage1_code_generation}
\label{subsec:stage2_initial_specification}

The first methodological component jointly shapes the implementation and specification before the repair loop begins. VeriGen-C constrains both artifacts with verification knowledge encoded as fixed policies before the reported runs. Under these constraints, the C code exposes proof anchors and the ACSL expresses requirement-bearing behavior through those anchors using supported, automation-friendly forms. The ablations evaluate the effect of removing these policies under the same task interface.

\subsubsection{Verifier-Aware Implementation Guidance}
\label{subsubsec:stage1_knowledge_guidance}

Given requirement \(r\), the implementation generator produces \(c_0=\mathcal{G}_{code}(r)\). Its goal is to produce code that supports proof, not only code that looks functionally correct. \textbf{Challenge.} Ordinary C idioms often hide the state that WP needs. Pointer-heavy updates hide processed prefixes, multiple exits hide variants and post-state cases, repeated library calls create additional memory and precondition obligations, and compact expressions leave no stable witness for ACSL to name. \textbf{Policy.} VeriGen-C favors explicit bounds, named counters or witnesses, explicit flags that represent algorithmic state, simple monotonic loops, low-aliasing memory updates, and output state that can be referenced from ACSL. These choices expose proof anchors for later loop invariants and contracts, avoiding specifications that must reconstruct behavior from complex expressions, pointer offsets, or repeated calls to library routines.

\subsubsection{Proof-Oriented Specification Guidance}
\label{subsubsec:stage2_knowledge_guidance}

Given \(r\) and \(c_0\), the specification synthesizer produces \(s_0=\mathcal{G}_{spec}(r,c_0)\). \textbf{Challenge.} A semantically plausible function contract can still be unprovable if the loops do not carry the induction facts needed by the postcondition. An overly abstract contract can also force WP to reason about witnesses, library semantics, or whole-buffer properties that are not exposed by the implementation. \textbf{Policy.} VeriGen-C constrains loop-level ACSL to provide induction support over the implementation's exposed anchors, including simple bounds, frame facts, variants, monotonic counters, flag-state relations, and processed-prefix bridge invariants. Function contracts are constrained to state observable requirement behavior through parameters, globals, and \texttt{\textbackslash result}, align each semantic clause with available proof anchors, and avoid unsupported or fragile ACSL forms.

These policies are realized through prompts, preflight, and verifier-guided repair. Because ACSL function contracts are caller-visible, clauses are anchored in parameters, globals, \texttt{\textbackslash result}, and output buffers---the channels WP can relate compositionally to the implementation.

After both phases, VeriGen-C applies ACSL normalization. The normalizer removes Markdown fences, adds standard headers, canonicalizes \texttt{\textbackslash result} and \texttt{\textbackslash old}, orders clauses, checks annotation placement, and routes unsupported helper definitions to repair. It preserves requirement content, and the supported construct set follows the ACSL and Frama-C/WP documentation~\cite{acsl_manual,framac_wp_manual}.

\subsection{Diagnosis-Routed Verification Repair}
\label{subsec:stage3_verification_diagnosis}

The second methodological component turns verifier feedback into controlled refinement. VeriGen-C first performs deterministic ACSL preflight checks. If preflight succeeds, it invokes Frama-C/WP and normalizes the result into \texttt{Pass\_$m$\_$n$}, \texttt{Fail\_$m$\_$n$}, \textsc{Invalid}, or \textsc{Unknown}. \textbf{Challenge.} A non-pass WP outcome does not identify a single repair target. The same result tag may reflect malformed ACSL, missing loop induction support, an over-strong postcondition, a frame error, a memory-model obligation, or implementation shape that blocks proof. \textbf{Policy.} VeriGen-C filters raw verifier output into relevant invalidity messages or failing proof-goal summaries. It then diagnoses the failure category and uses routing guards to select specification repair, conservative joint code-specification repair, or termination.

\begin{algorithm}[!t]
    \caption{Diagnosis-routed repair loop}
    \label{alg:diagnosis_routed_repair}
    \small
    \begin{tabularx}{\columnwidth}{@{}rX@{}}
    \toprule
    1 & \textbf{while} repair budget $B$ remains and $v_i \neq \textsc{Pass}$ \textbf{do} \\
    2 & \quad Normalize verifier result and filtered report $(v_i,\rho_i)$. \\
    3 & \quad Diagnose failure $d_i \leftarrow \mathcal{D}(r,x_i,v_i,\rho_i,\sigma_i)$. \\
    4 & \quad Select route $a_i \leftarrow \mathsf{Route}(d_i,\sigma_i)$ from \{\textsc{spec}, \textsc{joint}, \textsc{stop}\}. \\
    5 & \quad \textbf{if} $a_i=\textsc{spec}$ \textbf{then} run specification repair. \\
    6 & \quad \textbf{else if} $a_i=\textsc{joint}$ \textbf{then} run joint code-specification repair. \\
    7 & \quad \textbf{else} return terminal failure with the latest artifact and verifier report. \\
    8 & \quad Re-normalize the repaired artifact and rerun Frama-C/WP to obtain $v_{i+1}$. \\
    9 & \quad Update $\sigma_{i+1}$ with diagnosis, route, artifact, verifier result, and budget state. \\
    10 & \textbf{end while} \\
    11 & Return the latest artifact and verifier result. \\
    \bottomrule
    \end{tabularx}
\end{algorithm}

Algorithm~\ref{alg:diagnosis_routed_repair} isolates the repair loop. Passing artifacts leave this loop and enter the requirement-alignment check; non-pass artifacts are diagnosed and routed to specification repair, joint code-specification repair, or terminal failure.

\subsubsection{Diagnosis Knowledge Policy}
\label{subsubsec:stage3_diagnosis_knowledge_policy}

The diagnosis layer maps low-level verifier evidence into an actionable control signal:
\[
    \mathcal{D}(r,x_i,v_i,\rho_i,\sigma_i) \mapsto (f_i,a_i,e_i),
\]
where \(x_i=(c_i,s_i)\) is the current artifact, \(\rho_i\) is the filtered WP report, \(f_i\) is the failure category, \(a_i\) is the recommended action, and \(e_i\) is the supporting evidence. The policy separates \emph{what failed} from \emph{what should change next}. Invalid ACSL maps to specification repair, and unproved obligations map by default to specification-only proof repair. Joint repair becomes eligible only after the minimum specification-only attempts and persistent evidence that the implementation shape blocks proof, such as hidden witnesses, repeated string-library obligations, contract-invisible buffers, constant or ternary variants, or break-heavy control flow. The diagnosis may summarize root causes and repair targets, but deterministic failure categories and routing guards remain authoritative.

\subsubsection{Diagnosis-Conditioned Specification Repair}
\label{subsubsec:diagnosis_conditioned_spec_repair}

Specification repair keeps the implementation fixed and updates only the ACSL side:
\[
    (c_i,s_{i+1})=\mathcal{R}_{spec}(r,c_i,s_i,v_i,d_i).
\]
Specification repair fixes the ACSL side while preserving the implementation. For \textsc{Invalid} results, it makes the annotations acceptable to Frama-C again by repairing syntax, scoping, clause ordering, annotation placement, frames, variants, or unsupported constructs. For \texttt{Fail\_*} results, it targets the failing proof-obligation family by adjusting loop facts, frame clauses, postconditions, and brittle quantifier-heavy claims. When a proof failure involves an over-strong semantic clause, repair reshapes that clause around implementation-exposed anchors while preserving requirement-bearing content. A repaired artifact is counted as successful only if it later passes WP and is accepted by requirement-alignment checking.

Joint code and specification repair is deliberately conservative:
\[
    (c_{i+1},s_{i+1})=\mathcal{R}_{joint}(r,c_i,s_i,v_i,d_i).
\]
This route may update both the C implementation and the ACSL specification. It is never triggered by \textsc{Invalid} results and becomes eligible only after specification-only attempts and code-shape evidence. Code edits are allowed only to expose proof anchors or remove proof-hostile structures, such as hidden witnesses, repeated library calls, unbounded scans, or break-heavy control flow. After each repair, VeriGen-C reapplies extraction and ACSL normalization, then records the new artifact and verifier evidence in the task state.

\subsection{Requirement-Alignment Check}
\label{sec:reqalign-checking}

The third methodological component addresses the fact that WP success is specification-relative. Once WP reaches \textsc{Pass}, VeriGen-C sends the WP-passing artifact to the requirement-alignment check. \textbf{Challenge.} A verifier may prove a contract that is formally valid but too weak, for example one containing only safety, frame, nullability, range, or return-shape clauses. Counting such artifacts as successful would overstate verified-code generation. \textbf{Policy.} The checker accepts contracts with non-trivial requirement-bearing behavior and rejects shell-only, inconsistent, or unclear specifications. A \emph{shell-only} contract contains only safety, frame, nullability, range, or shape clauses and no task-specific functional \texttt{ensures} clause; \emph{safety-only} is the special case restricted to memory safety and undefined behavior. A contract carries \emph{non-trivial requirement-bearing behavior} when at least one functional clause relates an observable output---return value, modified parameter or global, or output buffer---to a stated input condition. An \emph{inconsistent} contract contradicts a requirement clause for the same valid input; an \emph{unclear} contract cannot be reliably matched. Accepted buckets are \emph{slice} (at least one branch or direction), \emph{core} (main input--output relation), and \emph{full} (all stated task-specific properties). Given the original requirement \(r\) and the final WP-passing specification \(s^\star\), the checker returns
\[
    \mathsf{ReqAlign}(r,s^\star)=\langle o,C,M,E\rangle,
\]
where \(o\) is the normalized outcome bucket, \(C\) is clause-level coverage evidence, \(M\) is the set of detected mismatches, and \(E\) is a textual explanation. The outcome distinguishes accepted \texttt{slice}, \texttt{core}, and \texttt{full} contracts from \texttt{shell-only}, \texttt{inconsistent}, or \texttt{unclear} contracts. The predicate \(\mathsf{Accept}(\mathsf{ReqAlign}(r,s^\star))\) holds when \(o\) belongs to the accepted bucket set used in the evaluation. Otherwise, the WP-passing artifact is rejected as not requirement-aligned.

Alignment has three steps: a deterministic prescreen rejects contracts without functional \texttt{ensures} or with only safety and frame properties; an LLM decomposes the requirement into clauses and returns covered, missing, contradictory, or ambiguous records with matching ACSL fragments; deterministic rules map these records to the outcome bucket. \(E\) records the audit rationale, while acceptance uses \(o\), \(C\), and \(M\). Frama-C/WP establishes proof; the checker establishes whether the proved contract retains task-specific content. A requirement is in scope when it names input objects, a result channel, and at least one task-specific rule; otherwise it is labeled unclear. Rejected artifacts retain their evidence and remain outside the repair loop.

\section{Evaluation}
\label{sec:evaluation}

We evaluate VeriGen-C along two complementary dimensions. The first is proof reachability: whether the generated C+ACSL artifact can be automatically proved by Frama-C/WP. The second is requirement alignment: whether the proved contract carries non-trivial requirement-level behavior rather than only safety or shape constraints. This distinction matters because WP proves conformance to the generated ACSL contract, not to the original natural-language requirement. A WP-passing artifact is therefore useful only when the verified contract preserves meaningful requirement intent.

Our evaluation addresses four research questions:

\begin{itemize}
    \item \textbf{RQ1: Overall effectiveness.}
    How effectively does VeriGen-C generate WP-verified requirement-aligned C+ACSL artifacts across benchmarks?

    \item \textbf{RQ2: Verification knowledge guidance.}
    How much does verification-oriented guidance contribute to verification success and requirement-aligned quality?

    \item \textbf{RQ3: Diagnosis and routed repair.}
    How important are verifier-driven diagnosis, failure classification, and routed repair for iterative refinement?

    \item \textbf{RQ4: Model sensitivity and cost.}
    How sensitive is VeriGen-C to the underlying language model, and what runtime and token costs does it incur?

\end{itemize}

RQ1 evaluates the full VeriGen-C workflow end to end. RQ2 and RQ3 then isolate the effects of verification knowledge guidance and diagnosis-routed repair, and RQ4 examines how model choice affects quality and cost. Across all four questions, requirement-alignment checking is applied to WP-passing artifacts so that formally verified but weak contracts are not counted as successful outputs.

\subsection{Experimental Setup}
\label{subsec:experimental_setup}

All methods use the same task interface, verifier configuration, requirement-alignment checker, and execution protocol. RQ2 and RQ3 vary workflow components, while RQ4 varies the model backend.

\noindent\textbf{VeriGen-C configuration.} The default run follows the workflow described in Section~\ref{sec:method}: initial C generation, ACSL synthesis, Frama-C/WP validation, verifier-driven diagnosis, and iterative repair. The repair budget allows at most 5 specification-only attempts followed, when route guards permit, by at most 3 joint code--specification attempts. Joint repair becomes eligible only after 2 specification-only attempts and when diagnosis indicates code-shape evidence. Requirement-alignment checking runs only after WP verification succeeds.

\noindent\textbf{Frama-C/WP configuration.} 
We use Frama-C/WP as the deductive verification backend for all ACSL-annotated C programs~\cite{kirchner2015framac,framac_wp_manual}. WP is configured with Alt-Ergo and Z3~\cite{altergo_official,demoura2008z3}, the \texttt{Typed+real} model, and an 8-second timeout per proof obligation. A task is counted as WP-passing only when all generated proof obligations are discharged under this fixed configuration. For each verification run, we record the final WP status and the numbers of proved and generated goals.

\noindent\textbf{Language models:} The main VeriGen-C results, the one-shot reference, and the controlled comparisons use the same GPT-5.4 with temperature 0.0. Performance differences are therefore attributable to workflow design rather than model choice. RQ4 studies model sensitivity by replacing the underlying model with GPT-4.1, Gemini, or DeepSeek.

\noindent\textbf{Compared methods.} We compare VeriGen-C with a direct \emph{one-shot} reference, an iterative \emph{No-KG} control, and a \emph{No-diagnose} ablation. The one-shot reference asks the same base model to generate a complete C+ACSL artifact in a single call and verifies it once. No-KG keeps the multi-stage workflow, repair limits, and task interface, but replaces verification knowledge guidance with generic task instructions. No-diagnose keeps the same initial code and specification generation as VeriGen-C but disables verifier-driven diagnosis and sends every non-pass artifact directly to joint repair. We report WP pass rate as the proof-reachability measure and WP-verified requirement-aligned success as the primary metric.

\noindent\textbf{Execution protocol.} Each benchmark instance is run as an independent task. Generation, verification, repair, and requirement-alignment checking are executed automatically, with no manual edits to generated artifacts. For each task, we record the final WP result, verification attempts, final non-pass category when applicable, requirement-alignment bucket when WP succeeds, token logs when available, and final task-level outcome. Intermediate artifacts and reports are saved so that each task outcome can be audited later.

\subsection{Datasets}
\label{subsec:datasets}

\begin{table}[t]
\centering
\caption{Benchmarks and retained-task construction used in the evaluation.}
\label{tab:dataset_summary}
\footnotesize
\setlength{\tabcolsep}{3pt}
\begin{tabularx}{\columnwidth}{lrX}
\toprule
Dataset & Tasks & Retained-task construction \\
\midrule
HumanEval-X & 164 & Used all HumanEval-X C++ prompts and normalized them to the common C/ACSL task interface. \\
LCB Lite & 201 & Retained 201 of 342 candidates that fit a deterministic single-function C/ACSL interface; rewrote platform-style statements with explicit parameters, caller-provided output buffers, and no stdin/stdout. \\
FM-Bench & 70 & Retained 70 of 90 function-level records from the FM-Bench test split; lightly normalized source-language descriptions to C scalar, array, struct, and enum interfaces. \\
\bottomrule
\end{tabularx}
\end{table}

We evaluate VeriGen-C on three benchmarks derived from HumanEval-X~\cite{zheng2023codegeex}, LiveCodeBench Lite~\cite{jain2024livecodebench}, and FM-Bench~\cite{cao-etal-2025-informal}.
These benchmarks cover different requirement styles and share a deterministic function-level C/ACSL setting after preprocessing.
Table~\ref{tab:dataset_summary} summarizes the retained tasks and preprocessing used in the evaluation.

We use a pre-specified interface filter. HumanEval-X contributes all 164 C++ prompts, normalized to C. FM-Bench contains 3,965 records, 90 of them function-level; we retain 70 and exclude 20 with Lean type or proof objects, metaprogramming, or generic and dependent-type frameworks. LiveCodeBench Lite contributes 201 of 342 after excluding batch I/O, numerical tolerance, graph/tree, and constructive or non-unique outputs. The filter was fixed before model runs; source IDs and exclusion reasons are recorded in the evaluation outputs.

\begin{table}[!t]
  \centering
  \caption{Validation evidence for the requirement-alignment checker.}
  \label{tab:reqalign_checker_validation}
  \footnotesize
  \setlength{\tabcolsep}{10pt}
  \begin{tabular}{lcc}
  \toprule
  Validation item & Coverage & Audit result \\
  \midrule
  Clause-coverage reports & 298/298 & Complete \\
  Multi-class category labels & 289/298 & $\kappa=0.96$ \\
  Binary accept/reject labels & 290/298 & $\kappa=0.91$ \\
  Rejected WP-passing artifacts & 51/298 & Audited \\
  \bottomrule
  \end{tabular}
\end{table}

\subsection{Metrics}
\label{subsec:metrics}

The result tables report WP reachability together with the stricter WP-verified requirement-aligned success metric.

\noindent\textbf{WP-verified requirement-aligned success.} Let $\mathcal{T}$ be the benchmark task set and $\mathcal{T}_{\mathrm{wp\_pass}}$ the tasks whose final WP result is \texttt{Pass\_*}. Requirement-alignment checking is applied only to $\mathcal{T}_{\mathrm{wp\_pass}}$. We accept \(\mathcal{B}_{\mathrm{acc}}=\{\texttt{slice},\texttt{core},\texttt{full}\}\), where the three labels denote increasing coverage of requirement-bearing behavior. We define WP-verified requirement-aligned success as the fraction of all benchmark tasks whose final artifacts both pass WP and fall into \(\mathcal{B}_{\mathrm{acc}}\). This metric measures verified non-vacuous requirement content, not complete requirement formalization. A slice-level accepted artifact is a non-vacuous verified specification slice, not a complete formalization of the requirement. The evaluation outputs record the corresponding internal checker buckets for reproducibility.

\noindent\textbf{WP reachability.} WP pass rate measures verifier reachability before requirement-alignment filtering. We define it as the fraction of all benchmark tasks whose final WP result is \texttt{Pass\_*}. We also report \texttt{Invalid}, \texttt{Fail}, and \texttt{Unknown}: \texttt{Invalid} denotes preflight or syntax rejection, \texttt{Fail} denotes unproved or timed-out obligations, and \texttt{Unknown} denotes an unclear or unnormalized verifier outcome.

\noindent\textbf{Requirement-alignment acceptance and rejection.} For each WP-passing artifact, the checker accepts \(\mathcal{B}_{\mathrm{acc}}\) and rejects \(\mathcal{B}_{\mathrm{rej}}=\{\texttt{shell-only},\texttt{inconsistent},\texttt{unclear}\}\); Section~\ref{sec:reqalign-checking} defines these buckets. In the result tables, Success denotes WP-verified requirement-aligned success. ReqAlign \(\Delta\) is Success Rate minus WP Rate; negative values indicate the percentage-point loss after requirement-alignment filtering among WP-passing artifacts. Shell-only WP reports the subset of WP-passing artifacts rejected because their contracts contain only shell-level properties.

\noindent\textbf{Auxiliary measures.} For repair analysis, we use saved outcome summaries to distinguish initial and repaired WP passes, first-pass iteration, and the route that first produces a final pass. For model-sensitivity experiments, we report token usage as the common cost measure available across model backends.

\subsection{Validation of the Requirement-Alignment Check}
\label{subsec:reqalign_checker_validation}

WP pass alone can over-count artifacts whose contracts are formally proved but semantically weak. We therefore audit requirement-alignment decisions on the main-run WP-passing artifacts. These 298 artifacts are the WP-passing outputs produced by the main VeriGen-C runs across the three benchmarks, so the audit evaluates the same artifacts used in the reported RQ results. The audit uses a clause-level rubric. One auditor with C/ACSL and Frama-C/WP experience labeled all 298 artifacts without seeing the checker outputs. Each requirement clause had to match an ACSL fragment; absent support was omission, and incompatible assertions under the same valid input were contradiction.

Table~\ref{tab:reqalign_checker_validation} summarizes the validation evidence. The checker covers all 298 WP-passing main-run artifacts and produces clause-level coverage reports for each, making every accept/reject decision auditable against the source requirement. The auditor and checker agreed on 289/298 category labels (97.0\%, $\kappa=0.96$) and 290/298 accept/reject decisions (97.3\%, $\kappa=0.91$). The disagreements comprised four contradictory contracts accepted by the checker, four slice-level contracts rejected, and one full contract labeled slice without changing acceptance. The study uses a single expert auditor; $\kappa$ therefore measures auditor--checker agreement. The 51 rejected WP-passing artifacts show why WP success alone would over-count weak contracts.

Overall, these results support using the checker as an evaluation filter because it is complete over WP-passing outputs, auditable at the clause level, and stable under manual audit.

\subsection{Experimental Results}
\noindent\textbf{(1) RQ1: Overall Effectiveness.}

\noindent\textbf{Experimental design.} RQ1 compares the full VeriGen-C workflow with the interface-compatible one-shot baseline on HumanEval-X, FM-Bench, and LiveCodeBench Lite. Both methods use the same task interface, base model, verifier configuration, and requirement-alignment checker. The primary metric is WP-verified requirement-aligned success; WP pass rate is reported separately to show proof reachability before requirement-alignment filtering.

\begin{table}[t]
\centering
\caption{RQ1: end-to-end effectiveness across datasets.}
\label{tab:rq1_main_results}
\footnotesize
\setlength{\tabcolsep}{5pt}
\begin{tabular}{llccc}
\toprule
Dataset & Method & WP Pass & Success & Success Gain \\
\midrule
HumanEval-X & VeriGen-C & 95/164 & \textbf{73/164} & +39.0 pp \\
HumanEval-X & One-shot  & 13/164 & 9/164  & -- \\
\midrule
FM-Bench & VeriGen-C & 61/70 & \textbf{56/70} & +35.7 pp \\
FM-Bench & One-shot  & 31/70 & 31/70 & -- \\
\midrule
LCB Lite & VeriGen-C & 142/201 & \textbf{118/201} & +52.2 pp \\
LCB Lite & One-shot  & 21/201  & 13/201  & -- \\
\bottomrule
\end{tabular}
\vspace{2pt}
\begin{flushleft}
\footnotesize
\emph{Note.} Success Gain is the percentage-point improvement in Success Rate over One-shot.
\end{flushleft}
\end{table}

\begin{table*}[t]
    \centering
    \caption{RQ2: Effect of verification knowledge guidance across datasets.}
    \label{tab:rq2_knowledge_main}
    \footnotesize
    \setlength{\tabcolsep}{12pt}
    \begin{tabular}{llrrrrrc@{\hspace{8pt}}c}
    \toprule
    Dataset & Method & Total & WP & WP Rate & Success & Success Rate & ReqAlign $\Delta$ & Shell-only WP \\
    \midrule
HumanEval-X & VeriGen-C & 164 & 95  & 57.9\% & 73  & \textbf{44.5\%} & -13.4 pp & 9/95 (9.5\%) \\
    HumanEval-X & No-KG     & 164 & 37  & 22.6\% & 22  & 13.4\% & -9.2 pp  & 14/37 (37.8\%) \\
    HumanEval-X & One-shot  & 164 & 13  & 7.9\%  & 9   & 5.5\%  & -2.4 pp  & 4/13 (30.8\%) \\
    \midrule
FM-Bench & VeriGen-C & 70 & 61 & 87.1\% & 56 & \textbf{80.0\%} & -7.1 pp  & 0/61 (0.0\%) \\
    FM-Bench & No-KG     & 70 & 42 & 60.0\% & 31 & 44.3\% & -15.7 pp & 8/42 (19.0\%) \\
    FM-Bench & One-shot  & 70 & 31 & 44.3\% & 31 & 44.3\% & 0.0 pp  & 0/31 (0.0\%) \\
    \midrule
LCB Lite & VeriGen-C & 201 & 142 & 70.6\% & 118 & \textbf{58.7\%} & -11.9 pp & 8/142 (5.6\%) \\
    LCB Lite & No-KG     & 201 & 104 & 51.7\% & 50  & 24.9\% & -26.8 pp & 29/104 (27.9\%) \\
    LCB Lite & One-shot  & 201 & 21  & 10.4\% & 13  & 6.5\%  & -3.9 pp  & 8/21 (38.1\%) \\
    \bottomrule
    \end{tabular}
\end{table*}

Table~\ref{tab:rq1_main_results} shows that VeriGen-C improves WP-verified requirement-aligned success on all three datasets: from 5.5\% to 44.5\% on HumanEval-X, from 44.3\% to 80.0\% on FM-Bench, and from 6.5\% to 58.7\% on LiveCodeBench Lite. These gains cover compact task-oriented prompts (HumanEval-X), specification-like function descriptions (FM-Bench), and platform-derived problem statements after I/O removal (LiveCodeBench Lite), indicating that the improvement is not tied to a single requirement style. The absolute success rates are also consistent with how close the inputs are to function-level specifications. FM-Bench obtains the highest success rate, consistent with its source descriptions already naming functions, data models, and branch behavior. HumanEval-X and LiveCodeBench Lite require more reconstruction of implicit behavior from examples or problem narratives, which leaves more room for missing contract clauses and mismatches that affect verification. WP pass rates also increase substantially, with gains of 50.0, 42.9, and 60.2 percentage points, showing that the workflow improves verifier reachability as well as requirement-aligned success.

The gap between WP pass and Success shows why WP pass alone would over-count successful artifacts. For example, 95 HumanEval-X artifacts pass WP, but only 73 are accepted after requirement-alignment checking; on LiveCodeBench Lite, the corresponding counts are 142 and 118. Persisted outcome summaries further show that the one-shot baseline ends with many front-end invalid outcomes: 142/164 on HumanEval-X, 33/70 on FM-Bench, and 132/201 on LiveCodeBench Lite. In contrast, VeriGen-C obtains 53, 13, and 89 WP passes after repair on the three datasets. This indicates that the end-to-end gains come not only from initial generation but also from successful recovery within the bounded workflow; RQ3 isolates the repair mechanism.

\begin{tcolorbox}[size=title]
\textit{\textbf{Answer to RQ1:} VeriGen-C improves end-to-end verified-code generation on all three benchmarks, raising WP-verified requirement-aligned success by 35.7--52.2 percentage points over one-shot generation. The gains appear at both evaluation boundaries: more artifacts pass WP, and more WP-passing artifacts preserve non-trivial requirement behavior in their contracts. These results support joint construction of verifier-friendly implementations and ACSL specifications.}
\end{tcolorbox}

\noindent\textbf{(2) RQ2: Effect of Verification Knowledge Guidance.}

\noindent\textbf{Experimental design.} RQ2 compares VeriGen-C, No-KG, and One-shot under the same task interface, model, verifier configuration, and requirement-alignment checker. The comparison between VeriGen-C and No-KG isolates verification knowledge guidance within the same multi-stage workflow; the comparison between No-KG and One-shot separates workflow decomposition from domain-specific guidance.

No-KG uses 10.84/8.06/9.56 calls and 4.42/3.03/3.78 iterations versus 10.21/6.29/9.50 and 4.10/2.14/3.75 for VeriGen-C, yet obtains 37/42/104 WP passes versus 95/61/142. On HumanEval-X, token use is 26.7K versus 51.0K; we report calls, iterations, and tokens for budget transparency.

Table~\ref{tab:rq2_knowledge_main} shows that removing verification knowledge guidance substantially reduces WP-verified requirement-aligned success under the same workflow. VeriGen-C outperforms No-KG by 31.1, 35.7, and 33.8 percentage points on HumanEval-X, FM-Bench, and LiveCodeBench Lite, respectively. No-KG often improves WP reachability over One-shot, but the extra WP-passing artifacts do not reliably become Success cases: on FM-Bench, No-KG reaches 42 WP-passing artifacts but only 31 Success cases, matching One-shot; on LiveCodeBench Lite, 54 of its 104 WP-passing artifacts are filtered out at the requirement-alignment boundary.

The shell-only WP column explains this gap. Removing knowledge guidance increases shell-only WP artifacts from 9.5\% to 37.8\% on HumanEval-X, from 0.0\% to 19.0\% on FM-Bench, and from 5.6\% to 27.9\% on LiveCodeBench Lite. Thus, staged generation can improve WP reachability, but without verification knowledge guidance it does not reliably preserve the functional clauses that express the original requirement. RQ3 separately isolates repair routing.

\begin{tcolorbox}[size=title]
\textit{\textbf{Answer to RQ2:} Verification knowledge guidance is necessary for turning staged generation into requirement-aligned verified artifacts. Removing it lowers Success Rate by 31.1--35.7 percentage points and increases shell-only WP artifacts across all three datasets. This shows that workflow decomposition alone can improve proof reachability, but verification-specific constraints are needed to preserve proof anchors, supported ACSL forms, and functional requirement clauses.}
\end{tcolorbox}

\noindent\textbf{(3) RQ3: Effect of Diagnosis and Routed Repair.}

\noindent\textbf{Experimental design.} RQ3 compares VeriGen-C with a No-diagnose ablation under the same model, verifier configuration, workflow budget, and initial code/specification policies. VeriGen-C diagnoses each non-pass result and routes repair between specification-only and joint code--specification revision; No-diagnose disables diagnosis and sends every non-pass result directly to joint repair.

\begin{table*}[t]
    \centering
    \caption{RQ3: Effect of verifier-driven diagnosis and routed repair across datasets.}
    \label{tab:rq3_diagnosis_main}
    \footnotesize
    \setlength{\tabcolsep}{8.5pt}
    \begin{tabular}{lllrrrrrc@{\hspace{8pt}}c}
    \toprule
    Dataset & Method & Repair strategy & Total & WP & WP Rate & Success & Success Rate & ReqAlign $\Delta$ & Shell-only WP \\
    \midrule
HumanEval-X & VeriGen-C & Spec$\rightarrow$joint & 164 & 95 & 57.9\% & 73 & \textbf{44.5\%} & -13.4 pp & 9/95 (9.5\%) \\
    HumanEval-X & No-diagn. & Direct joint & 164 & 80 & 48.8\% & 39 & 23.8\% & -25.0 pp & 36/80 (45.0\%) \\
    \midrule
FM-Bench & VeriGen-C & Spec$\rightarrow$joint & 70 & 61 & 87.1\% & 56 & \textbf{80.0\%} & -7.1 pp  & 0/61 (0.0\%) \\
    FM-Bench & No-diagn. & Direct joint & 70 & 58 & 82.9\% & 9  & 12.9\% & -70.0 pp & 46/58 (79.3\%) \\
    \midrule
LCB Lite & VeriGen-C & Spec$\rightarrow$joint & 201 & 142 & 70.6\% & 118 & \textbf{58.7\%} & -11.9 pp & 8/142 (5.6\%) \\
    LCB Lite & No-diagn. & Direct joint & 201 & 128 & 63.7\% & 54  & 26.9\% & -36.8 pp & 58/128 (45.3\%) \\
    \bottomrule
    \end{tabular}
\end{table*}

Table~\ref{tab:rq3_diagnosis_main} shows that diagnosis and routed repair mainly affect the quality of WP-passing artifacts. The WP-rate gaps between VeriGen-C and No-diagnose are modest, but the Success-rate gaps are much larger: 20.7, 67.1, and 31.8 percentage points on HumanEval-X, FM-Bench, and LiveCodeBench Lite. The shell-only column explains the discrepancy: No-diagnose raises shell-only WP artifacts to 45.0\%, 79.3\%, and 45.3\% on the three datasets. FM-Bench is clearest: No-diagnose obtains 58 WP-passing artifacts, close to VeriGen-C's 61, but only 9 remain requirement-aligned.

These results show why direct joint repair is not a substitute for diagnosis. Broadly rewriting code and specification can remove some verifier-front-end errors, but it also increases the chance of satisfying WP with shell-level contracts. VeriGen-C keeps repair specification-only when the evidence points to ACSL surface issues or proof-obligation shape, and escalates to joint code--specification repair only when repeated evidence indicates code-shape blockers. Together with RQ2, these results show that verification knowledge and repair routing address distinct failure modes.

\begin{tcolorbox}[size=title]
\textit{\textbf{Answer to RQ3:} Diagnosis and routed repair are necessary for preserving specification quality during iteration. Direct joint repair can remove some verifier failures, but it often changes the artifact too broadly and increases shell-level WP passes. Routing repair by failure evidence keeps specification-only fixes separate from code-shape revisions, which is critical for improving proof outcomes without weakening the contract that WP proves.}
\end{tcolorbox}

\noindent\textbf{(4) RQ4: Model Sensitivity and Cost.}

\noindent\textbf{Experimental design.} RQ4 runs the same VeriGen-C workflow on HumanEval-X with GPT-4.1, GPT-5.4, Gemini 3.1 Pro Preview, and DeepSeek V4 Pro. The verifier configuration, repair budget, task set, and checker are fixed; only the backend changes. RQ1--RQ3 establish cross-dataset effects; RQ4 isolates model sensitivity and cost on 164 HumanEval-X tasks under the same verifier and repair budget.

\begin{table*}[t]
    \centering
    \caption{RQ4: Model sensitivity of VeriGen-C on HumanEval-X.}
    \label{tab:rq4_model_main}
    \footnotesize
    \setlength{\tabcolsep}{12pt}
    \begin{tabular}{lrrrrr@{\hspace{8pt}}cr}
    \toprule
    Model & WP & WP Rate & Success & Success Rate & ReqAlign $\Delta$ & Shell-only WP & Avg. Tok. \\
    \midrule
    GPT-4.1 & 77  & 47.0\% & 53 & 32.3\% & -14.7 pp & 24/77 (31.2\%) & 50.3K \\
    GPT-5.4 & 95  & 57.9\% & 73 & 44.5\% & -13.4 pp & 9/95 (9.5\%) & 51.0K \\
Gemini 3.1 Pro Preview & 124 & \textbf{75.6\%} & 77 & \textbf{47.0\%} & -28.6 pp & 47/124 (37.9\%) & 74.9K \\
    DeepSeek V4 Pro & 93 & 56.7\% & 60 & 36.6\% & -20.1 pp & 32/93 (34.4\%) & 91.3K \\
    \bottomrule
    \end{tabular}
    \vspace{2pt}
    \begin{flushleft}
    \footnotesize
    \emph{Note.} All models are evaluated on 164 HumanEval-X tasks using the same VeriGen-C workflow and verifier configuration. ``Avg. Tok.'' is average total tokens per task.
    \end{flushleft}
\end{table*}

Table~\ref{tab:rq4_model_main} shows that model choice affects both proof reachability and requirement-aligned quality. Within the GPT family, GPT-5.4 improves over GPT-4.1 from 47.0\% to 57.9\% in WP pass rate and from 32.3\% to 44.5\% in Success Rate, with comparable token use. Cross-provider results are non-monotonic: Gemini reaches the highest WP pass rate, 75.6\%, and the most Success cases, 77, but its shell-only WP share is 37.9\%; DeepSeek nearly matches GPT-5.4 in WP reachability but has lower Success Rate and the highest token use. Thus, model comparison must report WP reachability, requirement-aligned quality, and cost together.

The framework effect remains larger than a simple model-strength effect. Even with GPT-4.1, VeriGen-C reaches 47.0\% WP pass rate and 32.3\% Success Rate on HumanEval-X, substantially above the GPT-5.4 one-shot baseline in RQ1.

\begin{tcolorbox}[size=title]
\textit{\textbf{Answer to RQ4:} VeriGen-C is sensitive to the underlying model, but model choice does not replace the framework. Different backends shift the trade-off among WP reachability, requirement alignment, and cost: a model can produce more WP-passing artifacts while also producing more shell-level contracts. Evaluating verified-code generation therefore requires reporting proof success, alignment quality, and cost together rather than ranking models by WP pass rate alone.}
\end{tcolorbox}

\section{Discussion}
\label{sec:discussion}

\noindent\textbf{Implications.}
The evaluation suggests that natural-language-to-verified-C generation is constrained by the joint interface among C, ACSL, and Frama-C/WP, not only by functional code generation. A successful artifact must be valid C/ACSL, expose proof-oriented program state, use verifier-supported annotation forms, and retain requirement-level specification content. One-shot often fails before meaningful proof search, and the No-KG control shows that staging alone is insufficient. Verification knowledge guidance shapes both code and specification toward forms that Frama-C/WP can parse, normalize, and prove while keeping code-level proof anchors aligned with specification-level requirement facts.

Repair must be routed by verifier evidence. A failed WP run may reflect malformed annotations, missing loop invariants, unsupported proof structure, an over-strong postcondition, or implementation shape that hides proof-relevant state. Treating every failure as a reason for joint regeneration can improve verifier acceptability while weakening the contract. VeriGen-C separates specification repair from joint code--specification revision and escalates only when the evidence indicates code-shape blockers. The RQ3 results show that repair improves verified-code generation only when it is constrained not to erase functional contract content.

\noindent\textbf{Interpreting WP-verified requirement-aligned success.}
\label{sec:reqalign-success-discussion}
WP proves that the C implementation satisfies the generated ACSL contract; it does not prove that the contract fully formalizes the natural-language requirement. We therefore count success only when an artifact both passes WP and is accepted by requirement-alignment checking. This criterion rejects shell-only and inconsistent contracts even when they are easy to prove. Accepted slice-level artifacts are verified non-vacuous specification slices, not complete human-level formalizations. The validation evidence in Section~\ref{subsec:reqalign_checker_validation} supports the checker as an evaluation filter.

\noindent\textbf{Verifier, benchmark, and model scope.}
\label{sec:wp-limitations}
The results are bounded by what Frama-C/WP can prove automatically. Proof success depends on contract shape, loop annotations, memory and arithmetic models, quantifier structure, and proving budget~\cite{framac_wp_manual,kirchner2015framac}. The benchmarks are function-level by design: tasks are normalized to C functions over scalars, arrays, strings, explicit lengths, output buffers, and small structures. This setting is appropriate for studying requirement-originated C/ACSL artifact generation, but it limits claims about larger C systems, concurrency, floating-point kernels, complex heap-manipulating code, and interactive I/O. Model choice changes proof reachability, requirement alignment, and cost, so VeriGen-C is best viewed as an offline verified-code generation framework.

\noindent\textbf{Threats to validity.}
\label{sec:threats}
\textbf{Internal validity.} The controlled comparisons use the same verifier configuration, timeout budget, preprocessing pipeline, and batch runner, but implementation details such as ACSL normalization, output extraction, preflight checks, and failure parsing could affect exact counts. We mitigate this risk by persisting final artifacts, verification reports, diagnosis reports, requirement-alignment reports, and batch summaries for audit. The ablations isolate the two main mechanisms under a fixed interface rather than comparing unrelated implementations.

\textbf{Construct validity.} WP-verified requirement-aligned success is stricter than WP pass alone, but requirement-alignment acceptance is not a mechanically proved equivalence between natural language and ACSL. The checker may miss subtle omissions, penalize conservative contracts, or reflect assumptions in its policy. We reduce this risk by rejecting shell-only, inconsistent, and unclear specifications and by validating the checker with clause-coverage evidence and manual audit. Section~\ref{subsec:reqalign_checker_validation} reports high manual-audit agreement on the main-run WP-passing artifacts, but the audit was conducted by a single auditor.

\textbf{External validity.} The datasets cover three sources and multiple requirement styles, but they are normalized into function-level C tasks. This selection may under-represent larger programs, pointer-rich libraries, floating-point code, concurrency, and APIs requiring substantial environment models. We do not claim a ranking against tools with different native interfaces, such as existing-program specification generation, temporal-logic formalization, or proof repair in another language.

\textbf{Model and reproducibility validity.} The main experiments use one default model, and the model-sensitivity study covers the backends for which complete local outputs are available. Provider-side updates, decoding behavior, latency, and context handling may affect future runs. We mitigate this by using fixed verifier settings, persisting run artifacts and requirement-alignment evidence, and reporting cross-provider results under the same metrics.

\section{Conclusion}
This paper studied natural-language-to-verified-C generation as a joint code--specification generation problem in which proof reachability and requirement alignment must be achieved together. VeriGen-C addresses this problem by constructing C code and ACSL specifications around shared proof anchors, routing repair with verifier evidence, and filtering WP-passing artifacts through requirement-alignment checking. Across HumanEval-X, FM-Bench, and LiveCodeBench Lite, it substantially improves WP-verified requirement-aligned success over one-shot C+ACSL generation. The ablations show that verification knowledge and diagnosis-routed repair help preserve requirement alignment, while the model-sensitivity study shows that backend choice shifts proof, alignment, and cost trade-offs without replacing the framework.

\bibliographystyle{IEEEtran}
\bibliography{references}

@misc{stackoverflow2025devsurvey_ai,
  title        = {2025 Developer Survey: {Artificial Intelligence}},
  author       = {{Stack Overflow}},
  year         = {2025},
  howpublished = {\url{https://survey.stackoverflow.co/2025/ai}},
  note         = {Accessed 2026-06-29}
}

@misc{alphabet2024q3_earnings_call,
  title        = {Alphabet {Q3} 2024 Earnings Call},
  author       = {{Alphabet Inc.}},
  year         = {2024},
  howpublished = {\url{https://abc.xyz/2024-q3-earnings-call/}},
  note         = {Accessed 2026-06-29}
}

@manual{acsl_manual,
  title        = {{ACSL}: {ANSI/ISO C} Specification Language},
  author       = {Baudin, Patrick and Filli{\^a}tre, Jean-Christophe and March{\'e}, Claude and Monate, Benjamin and Moy, Yannick and Prevosto, Virgile},
  year         = {2021},
  url          = {https://frama-c.com/html/acsl.html},
  note         = {Accessed 2026-06-29}
}

@manual{framac_wp_manual,
  title        = {Frama-C WP Plug-in Manual},
  organization = {Frama-C},
  year         = {2024},
  url          = {https://frama-c.com/download/frama-c-wp-manual.pdf},
  note         = {Accessed 2026-06-29}
}

@inproceedings{cao-etal-2025-informal,
    title = "From Informal to Formal {--} Incorporating and Evaluating {LLM}s on Natural Language Requirements to Verifiable Formal Proofs",
    author = "Cao, Jialun  and
      Lu, Yaojie  and
      Li, Meiziniu  and
      Ma, Haoyang  and
      Li, Haokun  and
      He, Mengda  and
      Wen, Cheng  and
      Sun, Le  and
      Zhang, Hongyu  and
      Qin, Shengchao  and
      Cheung, Shing-Chi  and
      Tian, Cong",
    booktitle = "Proceedings of the 63rd Annual Meeting of the Association for Computational Linguistics (Volume 1: Long Papers)",
    month = jul,
    year = "2025",
    address = "Vienna, Austria",
    publisher = "Association for Computational Linguistics",
    url = "https://aclanthology.org/2025.acl-long.1310/",
    doi = "10.18653/v1/2025.acl-long.1310",
    pages = "26984--27003"
}

@misc{councilman2025formalverificationllmgeneratedcode,
      title={Towards Formal Verification of {LLM}-Generated Code from Natural Language Prompts}, 
      author={Aaron Councilman and David Jiahao Fu and Aryan Gupta and Chengxiao Wang and David Grove and Yu-Xiong Wang and Vikram Adve},
      year={2025},
      eprint={2507.13290},
      archivePrefix={arXiv},
      primaryClass={cs.PL},
      url={https://arxiv.org/abs/2507.13290}, 
}

@article{SU2026115177,
title = {{NL2ACSL}: Interactively Translating Natural Language to {ANSI C} Specification Language with Large Language Models},
journal = {Knowledge-Based Systems},
volume = {335},
pages = {115177},
year = {2026},
issn = {0950-7051},
doi = {10.1016/j.knosys.2025.115177},
url = {https://www.sciencedirect.com/science/article/pii/S0950705125022117},
author = {Wenxian Su and Xi Wu and Yongxin Zhao}
}

@misc{beg2026evaluatingllmgeneratedacslannotations,
      title={Evaluating {LLM}-Generated {ACSL} Annotations for Formal Verification}, 
      author={Arshad Beg and Diarmuid O'Donoghue and Rosemary Monahan},
      year={2026},
      eprint={2602.13851},
      archivePrefix={arXiv},
      primaryClass={cs.SE},
      url={https://arxiv.org/abs/2602.13851}, 
}

@article{wang2026preguss,
      title={A Tale of 1001 {LoC}: Potential Runtime Error-Guided Specification Synthesis for Verifying Large-Scale Programs},
      author={Zhongyi Wang and Tengjie Lin and Mingshuai Chen and Haokun Li and Mingqi Yang and Xiao Yi and Shengchao Qin and Yixing Luo and Xiaofeng Li and Bin Gu and Liqiang Lu and Jianwei Yin},
      journal={Proceedings of the ACM on Programming Languages},
      volume={10},
      number={OOPSLA1},
      pages={1874--1902},
      year={2026},
      publisher={ACM},
      doi={10.1145/3798268},
      url={https://doi.org/10.1145/3798268},
}

@INPROCEEDINGS{kozlenkov2002design,
  author={Kozlenkov, A. and Zisman, A.},
  booktitle={Proceedings IEEE Joint International Conference on Requirements Engineering}, 
  title={Are their design specifications consistent with our requirements?}, 
  year={2002},
  pages={145--154},
  doi={10.1109/ICRE.2002.1048517}
}

@inproceedings{ghosh2016arsenal,
  title     = {{ARSENAL}: Automatic Requirements Specification Extraction from Natural Language},
  author    = {Ghosh, Shalini and Elenius, Daniel and Li, Wenchao and Lincoln, Patrick and Shankar, Natarajan and Steiner, Wilfried},
  booktitle = {NASA Formal Methods},
  series    = {Lecture Notes in Computer Science},
  volume    = {9690},
  pages     = {41--46},
  year      = {2016},
  publisher = {Springer},
  doi       = {10.1007/978-3-319-40648-0_4}
}

@inproceedings{giannakopoulou2020fret,
  title     = {Formal Requirements Elicitation with {FRET}},
  author    = {Giannakopoulou, Dimitra and Mavridou, Anastasia and Rhein, Julian and Pressburger, Thomas and Schumann, Johann and Shi, Nija},
  booktitle = {Requirements Engineering: Foundation for Software Quality},
  series    = {Lecture Notes in Computer Science},
  volume    = {12045},
  pages     = {252--267},
  year      = {2020},
  publisher = {Springer},
  doi       = {10.1007/978-3-030-44429-7_18}
}

@article{giannakopoulou2021automated,
title = {Automated Formalization of Structured Natural Language Requirements},
journal = {Information and Software Technology},
volume = {137},
pages = {106590},
year = {2021},
issn = {0950-5849},
doi = {10.1016/j.infsof.2021.106590},
url = {https://www.sciencedirect.com/science/article/pii/S0950584921000707},
author = {Dimitra Giannakopoulou and Thomas Pressburger and Anastasia Mavridou and Johann Schumann}
}

@misc{hahn2022formal,
      title={Formal Specifications from Natural Language}, 
      author={Christopher Hahn and Frederik Schmitt and Julia J. Tillman and Niklas Metzger and Julian Siber and Bernd Finkbeiner},
      year={2022},
      eprint={2206.01962},
      archivePrefix={arXiv},
      primaryClass={cs.SE},
      url={https://arxiv.org/abs/2206.01962}, 
}

@InProceedings{cosler2023nl2spec,
author="Cosler, Matthias
and Hahn, Christopher
and Mendoza, Daniel
and Schmitt, Frederik
and Trippel, Caroline",
editor="Enea, Constantin
and Lal, Akash",
title="{nl2spec}: Interactively Translating Unstructured Natural Language to Temporal Logics with Large Language Models",
booktitle="Computer Aided Verification",
series="Lecture Notes in Computer Science",
volume="13965",
year="2023",
publisher="Springer Nature Switzerland",
address="Cham",
pages="383--396",
doi="10.1007/978-3-031-37703-7_18"
}

@INPROCEEDINGS{ma2025specgen,
  author={Ma, Lezhi and Liu, Shangqing and Li, Yi and Xie, Xiaofei and Bu, Lei},
  booktitle={2025 IEEE/ACM 47th International Conference on Software Engineering (ICSE)}, 
  title={{SpecGen}: Automated Generation of Formal Program Specifications via Large Language Models}, 
  year={2025},
  pages={16-28},
  doi={10.1109/ICSE55347.2025.00129}
}

@inproceedings{wen2024autospec,
  title={Enchanting Program Specification Synthesis by Large Language Models Using Static Analysis and Program Verification},
  author={Wen, Cheng and Cao, Jialun and Su, Jie and Xu, Zhiwu and Qin, Shengchao and He, Mengda and Li, Haokun and Cheung, Shing-Chi and Tian, Cong},
  booktitle={Computer Aided Verification},
  series={Lecture Notes in Computer Science},
  volume={14682},
  pages={302--328},
  year={2024},
  publisher={Springer Nature Switzerland},
  doi={10.1007/978-3-031-65630-9_16},
  url={https://doi.org/10.1007/978-3-031-65630-9_16}
}

@inproceedings{ma2025req2ltl,
  title={Bridging Natural Language and Formal Specification: Automated Translation of Software Requirements to {LTL} via Hierarchical Semantics Decomposition Using {LLM}s},
  author={Zhi Ma and Cheng Wen and Zhexin Su and Xiao Liang and Cong Tian and Shengchao Qin and Mengfei Yang},
  booktitle={Proceedings of the 40th IEEE/ACM International Conference on Automated Software Engineering},
  pages={1327--1339},
  year={2025},
  publisher={ACM},
  doi={10.1109/ASE63991.2025.00104},
  url={https://arxiv.org/abs/2512.17334},
  eprint={2512.17334},
  archivePrefix={arXiv},
  primaryClass={cs.SE}
}

@article{kirchner2015framac,
  author  = {Florent Kirchner and Nikolai Kosmatov and Virgile Prevosto and Julien Signoles and Boris Yakobowski},
  title   = {{Frama-C}: A Software Analysis Perspective},
  journal = {Formal Aspects of Computing},
  volume  = {27},
  number  = {3},
  pages   = {573--609},
  year    = {2015},
  doi     = {10.1007/s00165-014-0326-7}
}

@misc{ding2026fmagent,
      title={{FM-Agent}: Scaling Formal Methods to Large Systems via {LLM}-Based Hoare-Style Reasoning}, 
      author={Haoran Ding and Zhaoguo Wang and Haibo Chen},
      year={2026},
      eprint={2604.11556},
      archivePrefix={arXiv},
      primaryClass={cs.SE},
      url={https://arxiv.org/abs/2604.11556}, 
}

@misc{ma2026specsyn,
      title={{SpecSyn}: {LLM}-Based Synthesis and Refinement of Formal Specifications for Real-World Program Verification},
      author={Lezhi Ma and Shangqing Liu and Yi Li and Qiong Wu and Han Wang and Lei Bu},
      year={2026},
      eprint={2604.21570},
      archivePrefix={arXiv},
      primaryClass={cs.SE},
      url={https://arxiv.org/abs/2604.21570},
}

@inproceedings{li2026osvbench,
      title={{OSVBench}: Benchmarking {LLM}s on Specification Generation Tasks for Operating System Verification},
      author={Shangyu Li and Juyong Jiang and Tiancheng Zhao and Jiasi Shen},
      booktitle={Proceedings of the AAAI Conference on Artificial Intelligence},
      year={2026},
      volume={40},
      pages={31698--31707},
      doi={10.1609/aaai.v40i37.40437},
      url={https://doi.org/10.1609/aaai.v40i37.40437},
}

@misc{cheng2026sysmobench,
      title={{SysMoBench}: Evaluating {AI} on Formally Modeling Complex Real-World Systems},
      author={Qian Cheng and Ruize Tang and Emilie Ma and Finn Hackett and Peiyang He and Yiming Su and Ivan Beschastnikh and Yu Huang and Xiaoxing Ma and Tianyin Xu},
      year={2025},
      eprint={2509.23130},
      archivePrefix={arXiv},
      primaryClass={cs.SE},
      url={https://arxiv.org/abs/2509.23130},
}

@article{loughridge2024dafnybench,
      title={{DafnyBench}: A Benchmark for Formal Software Verification},
      author={Chloe Loughridge and Qinyi Sun and Seth Ahrenbach and Federico Cassano and Chuyue Sun and Ying Sheng and Anish Mudide and Md Rakib Hossain Misu and Nada Amin and Max Tegmark},
      journal={Transactions on Machine Learning Research},
      year={2025},
      url={https://openreview.net/forum?id=yBgTVWccIx},
}

@article{yang2025autoverus,
      title={{AutoVerus}: Automated Proof Generation for Rust Code},
      author={Chenyuan Yang and Xuheng Li and Md Rakib Hossain Misu and Jianan Yao and Weidong Cui and Yeyun Gong and Chris Hawblitzel and Shuvendu Lahiri and Jay Lorch and Shuai Lu and Fan Yang and Ziqiao Zhou and Shan Lu},
      journal={Proceedings of the ACM on Programming Languages},
      volume={9},
      number={OOPSLA2},
      pages={3454--3482},
      year={2025},
      publisher={ACM},
      doi={10.1145/3763174},
      url={https://doi.org/10.1145/3763174},
}

@inproceedings{zheng2023codegeex,
  title     = {{CodeGeeX}: A Pre-Trained Model for Code Generation with Multilingual Benchmarking on {HumanEval-X}},
  author    = {Zheng, Qinkai and Xia, Xiao and Zou, Xu and Dong, Yuxiao and Wang, Shan and Xue, Yufei and Wang, Zihan and Shen, Lei and Wang, Andi and Li, Yang and Su, Teng and Yang, Zhilin and Tang, Jie},
  booktitle = {Proceedings of the 29th ACM SIGKDD Conference on Knowledge Discovery and Data Mining},
  pages     = {5673--5684},
  year      = {2023},
  publisher = {ACM},
  doi       = {10.1145/3580305.3599790}
}

@inproceedings{jain2024livecodebench,
  title   = {{LiveCodeBench}: Holistic and Contamination Free Evaluation of Large Language Models for Code},
  author  = {Jain, Naman and Han, King and Gu, Alex and Li, Wen-Ding and Yan, Fanjia and Zhang, Tianjun and Wang, Sida and Solar-Lezama, Armando and Sen, Koushik and Stoica, Ion},
  booktitle = {The Thirteenth International Conference on Learning Representations},
  year    = {2025},
  eprint  = {2403.07974},
  archivePrefix = {arXiv},
  primaryClass  = {cs.SE},
  url     = {https://proceedings.iclr.cc/paper_files/paper/2025/hash/94074dd5a072d28ff75a76dabed43767-Abstract-Conference.html}
}

@inproceedings{demoura2008z3,
  author    = {de Moura, Leonardo and Bj{\o}rner, Nikolaj},
  title     = {{Z3}: An Efficient {SMT} Solver},
  booktitle = {Tools and Algorithms for the Construction and Analysis of Systems},
  series    = {Lecture Notes in Computer Science},
  volume    = {4963},
  pages     = {337--340},
  year      = {2008},
  publisher = {Springer},
  doi       = {10.1007/978-3-540-78800-3_24}
}

@misc{altergo_official,
  title        = {{Alt-Ergo} {SMT} Solver},
  organization = {OCamlPro},
  year         = {2026},
  url          = {https://alt-ergo.ocamlpro.com/},
  note         = {Accessed 2026-06-29}
}

@inproceedings{flanagan2001houdini,
  author    = {Flanagan, Cormac and Leino, K. Rustan M.},
  title     = {Houdini, an Annotation Assistant for {ESC/Java}},
  booktitle = {Formal Methods for Increasing Software Productivity},
  series    = {Lecture Notes in Computer Science},
  volume    = {2021},
  pages     = {500--517},
  year      = {2001},
  publisher = {Springer},
  doi       = {10.1007/3-540-45251-6_29}
}

@article{ernst2007daikon,
  author  = {Ernst, Michael D. and Perkins, Jeff H. and Guo, Philip J. and McCamant, Stephen and Pacheco, Carlos and Tschantz, Matthew S. and Xiao, Chen},
  title   = {The Daikon System for Dynamic Detection of Likely Invariants},
  journal = {Science of Computer Programming},
  volume  = {69},
  number  = {1--3},
  pages   = {35--45},
  year    = {2007},
  doi     = {10.1016/j.scico.2007.01.015}
}

@book{myers2011art,
  author    = {Glenford J. Myers and Corey Sandler and Tom Badgett},
  title     = {The Art of Software Testing},
  edition   = {3},
  publisher = {John Wiley \& Sons},
  year      = {2011},
  isbn      = {978-1-118-03196-4}
}

@article{li2026performance_ai_code,
  author  = {Li, Shuang and Cheng, Yuntao and Chen, Jinfu and Xuan, Jifeng and He, Sen and Shang, Weiyi},
  title   = {Performance Analysis of {AI}-Generated Code: A Case Study of Copilot, Copilot Chat, {CodeLLaMA}, and {DeepSeek-Coder} Models},
  journal = {Empirical Software Engineering},
  volume  = {31},
  number  = {3},
  pages   = {62},
  year    = {2026},
  doi     = {10.1007/s10664-025-10776-1}
}

@article{mugnier2025laurel,
  author  = {Mugnier, Eric and Anaya Gonzalez, Emmanuel and Polikarpova, Nadia and Jhala, Ranjit and Zhou, Yuanyuan},
  title   = {Laurel: Unblocking Automated Verification with Large Language Models},
  journal = {Proceedings of the ACM on Programming Languages},
  volume  = {9},
  number  = {OOPSLA1},
  pages   = {1519--1545},
  year    = {2025},
  doi     = {10.1145/3720499}
}

@misc{silva2025dafny,
  author       = {Silva, {\'A}lvaro and Mendes, Alexandra and Martins, Ruben},
  title        = {Inferring multiple helper {Dafny} assertions with {LLMs}},
  year         = {2025},
  howpublished = {arXiv:2511.00125},
  url          = {https://arxiv.org/abs/2511.00125}
}

\end{document}